# High-Pressure $Na_3(N_2)_4$, $Ca_3(N_2)_4$, $Sr_3(N_2)_4$, and $Ba(N_2)_3$ Featuring Nitrogen Dimers with Non-Integer Charges and Anion-Driven Metallicity

Dominique Laniel[1*], Bjoern Winkler[2], Timofey Fedotenko[1], Alena Aslandukova[3], Andrey Aslandukov[1], Sebastian Vogel[4], Thomas Meier[3], Maxim Bykov[5], Stella Chariton[6], Konstantin Glazyrin[7], Victor Milman[8], Vitali Prakapenka[6], Wolfgang Schnick[4], Leonid Dubrovinsky[3], Natalia Dubrovinskaia[1,9].

**Affiliations:**

[1]Material Physics and Technology at Extreme Conditions, Laboratory of Crystallography, University of Bayreuth, 95440 Bayreuth, Germany.

[2]Institut für Geowissenschaften, Abteilung Kristallographie, Johann Wolfgang Goethe-Universität Frankfurt, Altenhöferallee 1, D-60438, Frankfurt am Main, Germany.

[3]Bayerisches Geoinstitut, University of Bayreuth, 95440 Bayreuth, Germany

[4]Department of Chemistry, University of Munich (LMU) Butenandtstrasse 5–13, 81377 Munich, Germany.

[5]Department of Mathematics, Howard University, Washington, DC 20059, United States; The Earth and Planets Laboratory, Carnegie Institution for Science, Washington, DC 20015, United States

[6]Center for Advanced Radiation Sources, University of Chicago, Chicago, Illinois 60637, United States

[7]Photon Science, Deutsches Elektronen-Synchrotron, Notkestrasse 85, 22607 Hamburg, Germany.

[8]Dassault Systèmes BIOVIA, CB4 0WN Cambridge, United Kingdom

[9]Department of Physics, Chemistry and Biology (IFM), Linköping University, SE-581 83, Linköping, Sweden

*Correspondence to: dominique.laniel@uni-bayreuth.de

**Abstract**

Charged nitrogen dimers $[N_2]^{x-}$ are ubiquitous in high-pressure binary metal-nitrogen systems. They are known to possess integer formal charges $x$ varying from one through four. Here, we present the investigation of the binary alkali- and alkaline earth metal-nitrogen systems, Na-N, Ca-N, Sr-N, Ba-N to 70 GPa. We report on compounds—$Na_3(N_2)_4$, $Ca_3(N_2)_4$, $Sr_3(N_2)_4$, and $Ba(N_2)_3$—featuring $[N_2]^{x-}$ units with paradigm-breaking non-integer charges, $x = 0.67$, 0.75 and 1.5. The metallic nature of all four compounds is deduced from *ab initio* calculations. The conduction electrons occupy the $\pi^*$ antibonding orbitals of the $[N_2]^{x-}$ dimers that results in anion-driven metallicity. Delocalization of these electrons over the $\pi^*$ antibonding states enables the non-integer electron count of the dinitrogen species. Anion-driven metallicity is expected to be found among a variety of compounds with homoatomic anions (*e.g.*, polynitrides, carbides, and oxides), with the conduction electrons playing a decisive role in their properties.

**Introduction**

Nitrogen, the main component of the atmosphere, is omnipresent. Still, the chemistry of nitrogen has long been thought to be very limited due to the extreme stability of triple-bonded molecular nitrogen. As a result, nitrogen species seemed to be constrained to the nitride ($N^{3-}$) and the azide ($N_3^-$) anions. In the last decades, tremendous efforts led to the discovery and bulk stabilization of novel homoatomic ions such as $N_5^+$ and $N_5^-$,[1,2] providing a glimpse into nitrogen's potential chemical richness. High pressure investigations greatly expanded on these results, exhibiting the formation of polynitrogen species ($[N_4]^{4-}$,[3] $[N_5^-]$,[4–6] $[N_4]_\infty^{2-}$,[3,7,8] and N-frameworks[3,9–11]).

Studies of the chemistry of nitrogen at high densities revealed the ubiquity of charged $[N_2]^{x-}$ dimers in binary nitrides.[12,13,22–28,14–21] These were first discovered in the form of diazenides, $[N_2]^{2-}$, in $Sr_8N_4[N_2]\cdot(e^-)_2$, $Sr_8N_4[N_2]_2$, $SrN_2$, and $BaN_2$, with N–N bond lengths of 1.22 Å, comparable to those in protonated diazene $N_2H_2$ (1.21–1.25 Å).[29–31] The family of compounds featuring charged dinitrogen species was later enlarged due to high-pressure synthesis of transition metals dinitrides ($OsN_2$, $IrN_2$, $PtN_2$, $TiN_2$)[20,23,24] displaying tetravalent $[N_2]^{4-}$ anions with N–N bond distances of ~1.4 Å, similar to those in hydrazine $N_2H_4$ (1.47 Å).[32] The $[N_2]^{4-}$ anion is isoelectronic with the peroxide, $[O_2]^{2-}$, and so was dubbed "pernitride". Further progress in the exploration of the chemistry of nitrides brought findings of numerous compounds containing $[N_2]^{x-}$ structural units ($Li_2N_2$, $LiN_2$, $FeN_2$, $CrN_2$, $RuN_2$, $Re(N_2)N_2$, $Li_2Ca_3[N_2]_3$, etc.)[12,22,39,25,28,33–38] with N-N distances varying in a rather wide range (1.15 to 1.35 Å). For all of these compounds, the formal charge of the $[N_2]^{x-}$ anion is an integer, with $x = 1, 2, 3$, or 4, with the $\pi^*$ antibonding states of the nitrogen dimers populated.[40] The paradigm of integer formal charges observed in $[N_2]^{x-}$ dimers was recently questioned with the pressure synthesis of $CuN_2$, featuring [N$_2$]$^{x-}$ units with $x$ stated to be in between 1 and 2, as well as Na$_3$[N$_2$]$_4$[41] with $x$ = 0.75.[12] As the crystal chemistry of nitrides and their physical properties (compressibility, hardness, conductivity, optical characteristics, etc.) strongly depend on the character of [N$_2$]$^{x-}$ anions,[28,36] further high-pressure studies of yet unexplored metal-nitrogen systems are of the utmost interest and importance.

Here, we present the results of high-pressure single-crystal X-ray diffraction experiments in laser-heated diamond anvil cells (DAC), which reveal the novel binary alkali- and alkaline earth-nitrogen compounds Na$_3$(N$_2$)$_4$, Ca$_3$(N$_2$)$_4$, Sr$_3$(N$_2$)$_4$, and Ba(N$_2$)$_3$. The assignment of formal charges to the [N$_2$]$^{x-}$ units in each of these compounds yields a non-integer value. We show that the formal charges of the [N$_2$]$^{x-}$ species correlate with their intramolecular bond lengths and the physical properties of the compounds bearing them. The mechanism enabling this remarkable phenomenon is explored through *ab-initio* calculations and linked to the anion-driven metallicity of these compounds.

**Results and Discussion**

Sodium, calcium, strontium, and barium azides were each loaded into a BX90 diamond anvil cell[42] along with molecular nitrogen—acting both as a reagent and a pressure transmitting medium. Double-sided YAG laser-heating of the compressed samples was performed at various pressures between 40 and 70 GPa, as described in the Supplementary Information. The polycrystalline samples were then characterized by single-crystal X-ray diffraction.[43] The Na$_3$(N$_2$)$_4$, Ca$_3$(N$_2$)$_4$, and Sr$_3$(N$_2$)$_4$ compounds were synthesized at pressures of 41, 53, and 47 GPa, respectively, and found to be isostructural. Their structures adopt the tetragonal symmetry (space group *I*4$_1$/*amd*, #141); the lattice parameters are provided in Table 1. As an example of their common structure type, which is further referred to by its Pearson symbol *tI*44, the structure of Sr$_3$(N$_2$)$_4$ is shown in Figure 1a-d. The full crystallographic details can be found in Tables S1-S3. This structure type has recently been reported for K$_3$(N$_2$)$_4$ (at 27 GPa)[44] as well as Na$_3$(N$_2$)$_4$ (at 28 GPa)[41] and described in detail. A noteworthy feature of all *tI*44 compounds is that two crystallographically distinct nitrogen atoms, N1 and N2, form the two dimers, N1-N1 and N2-N2, with intramolecular distances almost identical within the experimental error.[41,44] The dimers are found to be elongated compared to nitrogen molecules (N$_2$), where the N-N distance is of ca. 1.10 Å[45,46] in nitrogen gas or in high pressure solid molecular nitrogen allotropes (Table 1). Upon decompression, the diffraction signal from Ca$_3$(N$_2$)$_4$ and Sr$_3$(N$_2$)$_4$ could be observed down to 6 and 2 GPa, respectively. The information on the equations of state of the three *tI*44 compounds studied in the present work is provided in the Supplementary Information (Figures S1-S3).

The laser-heating of Ba(N$_3$)$_2$ and N$_2$ at pressures between 51 and 69 GPa resulted in the formation of a new compound, Ba(N$_2$)$_3$, that adopts a cubic structure (space group *I*-43*d*, #220) with the lattice parameter $a$ = 9.983(2) Å ($V$ = 994.8(3) Å$^3$) at 65 GPa (Figure 1e-h). Based on the structure solution and refinement (Table S4), this structure type will be referred to by its Pearson symbol *cI*112.

Only two Wyckoff positions are occupied: 16$c$, by the barium Ba1 atom, and two times 48$e$, by the two crystallographically distinct nitrogen atoms N1 and N2. As seen in Figure 1g, barium atoms are coordinated by twelve [N$_2$]$^{x-}$ dimers—six side-on and six head-on. In comparison, the cations in the *tI*44 structure have only eight and nine coordinating dimers. At 65 GPa, the length of N1-N2 dimers $d$(N1-N2) = 1.111(14) Å is, within experimental error, similar to the length of the N$_2$ molecules in molecular nitrogen at the same pressure.[45,46] Each dimer has four barium first neighbors which form a distorted tetrahedron (Figure 1h). Considering the center of mass of the [N$_2$]$^{x-}$ dimer rather than the two individual atoms, Ba(N$_2$)$_3$ is isostructural to α-Cu$_3$As, which is related to the A15 structure type.[47] Ba(N$_2$)$_3$ could be decompressed down to 12 GPa; at lower pressures no diffraction lines characteristic of this phase could be observed. As discussed in the Supplementary Information, the *cI*112 structure type is more favorable at higher pressures than the *tI*44 for the alkali- and alkaline earth-nitrogen systems investigated here (see Table S5).

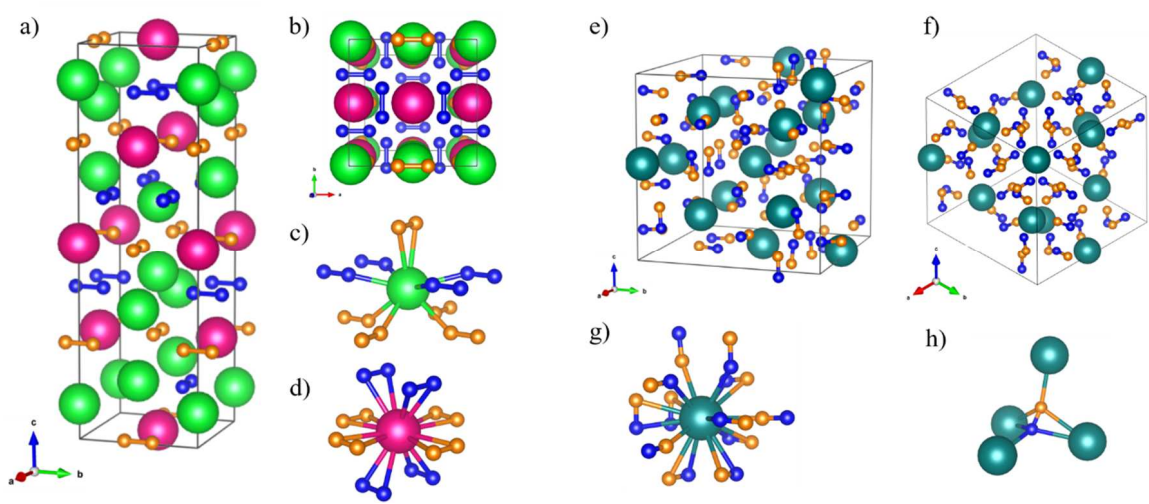

*Figure 1: Crystal structure of tI44 and cI112 compounds.* *a) tI44 Sr$_3$(N$_2$)$_4$ at 51 GPa; b) tI44 as viewed along the c-axis; c) Sr1 atom coordinated by nine [N$_2$] dimers, and d) Sr2 atom coordinated by eight dimers. Color code for spheres representing different atoms: green (Sr1), pink (Sr2), blue (N1), and orange (N2). The tI44 Sr$_3$(N$_2$)$_4$, Ca$_3$(N$_2$)$_4$, Na$_3$(N$_2$)$_4$ and K$_3$(N$_2$)$_4$ are isostructural. e) cI112 Ba(N$_2$)$_3$ at 65 GPa, f) cI112 as viewed along its three-fold axis; g) Ba coordinated by twelve dimers—six side-on and six head-on; h) The N1-N2 dimer coordinated by four Ba atoms. The color code for spheres representing different atoms: cyan (Ba1), blue (N1), and orange (N2).*

Density functional theory (DFT) calculations were performed for all three *tI*44 phases and the *cI*112 solid as detailed in the Supplementary Information. The relaxed structures were found to match with the experimental models, as seen in Tables S1-S5. The small decrease in the bond length (< 3%) of the calculated N-N distances in the [N$_2$]$^{x-}$ dimers between ambient pressure and 60 GPa (Table 1) highlights their incompressibility, in agreement with previous reports.[36,41]

*Table 1: Selected structural parameters of the metal-nitrogen tI44 and cI112 compounds. The full crystallographic details can be found in Tables S1-S4 of the Supplementary Information. The crystallographic data has been submitted under the deposition number CCDC 2080210-2080213.*

|  | Na$_3$(N$_2$)$_4$[41] | K$_3$(N$_2$)$_4$[44] | Ca$_3$(N$_2$)$_4$ | Sr$_3$(N$_2$)$_4$ | Ba(N$_2$)$_3$ |
|---|---|---|---|---|---|
| Pearson symbol | *tI*44 | *tI*44 | *tI*44 | *tI*44 | *cI*112 |
| Space group | *I4$_1$/amd* | *I4$_1$/amd* | *I4$_1$/amd* | *I4$_1$/amd* | *I-43d* |
| Pressure (GPa) | 28 | 27 | 53 | 47 | 65 |
| a (Å) | 4.9597(16) | 5.331(2) | 4.864(2) | 5.0747(11) | 9.983(2) |
| c (Å) | 16.29(7) | 17.552(6) | 15.758(6) | 16.343(4) | 9.983(2) |
| V (Å$^3$) | 400.7(18) | 498.8(5) | 372.8(3) | 420.88(17) | 994.8(3) |
| Exp. *d*(N-N) (Å) | 1.147(3) 1.149(3) | 1.151(11) 1.151(10) | 1.191(7) 1.194(6) | 1.19(4) 1.20(4) | 1.111(14) |
| Calc. *d*(N-N) (Å) (60 GPa) | 1.14 1.14 | 1.14 1.14 | 1.19 1.19 | 1.18 1.19 | 1.14 |
| Calc. *d*(N-N) (Å) (1 bar) | 1.15 1.16 | 1.15 1.16 | 1.20 1.22 | 1.20 1.22 | 1.15 |
| Formal charge of [N$_2$]$^{x-}$, *x* | 0.75 | 0.75 | 1.5 | 1.5 | 0.67 |

The formal charges $x$ = 0.75, 0.75, 1.5, 1.5, and 0.67, assigned to the [N$_2$]$^{x-}$ species in Na$_3$(N$_2$)$_4$, K$_3$(N$_2$)$_4$, Ca$_3$(N$_2$)$_4$, Sr$_3$(N$_2$)$_4$ and Ba(N$_2$)$_3$, respectively, are non-integer (Table 1), contrary to what could be expected for nitrogen compounds[12,20,23,25,48] and other solids featuring homoatomic species, namely carbon,[49–51] oxygen[52–54], and sulfur dimers.[55,56] This breaks the established paradigm of integer charges discussed above. The anion charges in the *tI*44 and *cI*112 solids were determined assuming that the cations of alkali and alkaline earth metals have formal charges equal to +1 and +2, respectively. At first glance, the validity of this hypothesis is not obvious, as for some subnitrides M$_2$N (M=Ca, Sr, Ba) the possibility of the cations' formal charge to be equal to +1.5 has been discussed.[29,57] However, arguments such as short interatomic metal-metal distances in Ca$_2$N, Sr$_2$N, and Ba$_2$N (12 to 18% shorter than in pure metals at ambient conditions), measurements of various physical properties (resistivity, photoelectron spectroscopy, compressibility, and others), and *ab initio* calculations, all advocate for subnitrides to be considered as metals described by the formula (M$^{2+}$)$_2$(N$^{3-}$)·$e^-$.[57,58]

As the non-integer formal charges of the [N$_2$]$^{x-}$ dimers in the *tI*44 and *cI*112 compounds are determined based on established crystallochemical considerations, it is more compelling to establish whether the value of *x* correlates with the geometry of the dimers and/or their chemical and physical properties. The first hint at a positive answer is that the [N$_2$]$^{x-}$ entities (see Table 1) are elongated in comparison with $N_2$ molecules in pure nitrogen at the same pressure.[45,46] Calculated intramolecular $[N_2]^{x-}$ bond lengths for numerous compounds at 1 bar are compared in Figure 2. Considering that calculated N-N intramolecular distances of 1.10 Å,[45,46] 1.16-1.20 Å,[12,13] 1.23-1.27 Å,[13,26] 1.30-1.34 Å [17–19] and 1.38-1.42 Å[17,20] are typical for $N_2$, $[N_2]^-$, $[N_2]^{2-}$, $[N_2]^{3-}$ and $[N_2]^{4-}$, respectively, those found in Table 1 for $Na_3(N_2)_4$, $K_3(N_2)_4$ and $Ba(N_2)_3$ are intermediate to $N_2$ and $[N_2]^-$ while those of $Ca_3(N_2)_4$ and $Sr_3(N_2)_4$ are in between of the usual values for $[N_2]^-$ and $[N_2]^{2-}$. As seen in Figure 2, there is a remarkable linear correlation between the N-N bond lengths and the formal charges of the $[N_2]^{x-}$ dimers. One can note that if integer charges were assigned to dimers in the *tI*44 and *cI*122 solids, they would have shown a significant deviation from a linear dependence. Further arguments in favor of the definite correlation between the dimers' non-integer formal charges and their chemical and physical characteristics are the behavior of the N-N vibrational modes and the compressibility of the $[N_2]^{x-}$-bearing compounds with respect to the *x* value. Indeed, the comparison of the calculated phonon density of ε-$N_2$, $Ba(N_2)_3$, $Ba_3(N_2)_4$ (a hypothetical compound) and $Ba(N_2)$ at 30 GPa in Figure S5 demonstrates a systematic redshift of the $[N_2]^{x-}$ stretching mode with increasing values of *x*. Also, Figure S6 shows a clear correlation between the compressibility of 15 reported binary compounds[16,22–24,59] containing $[N_2]^{x-}$ units and the value of $<d>^3/Z_{[M]}\cdot x$, where $<d>$ is the average cation-anion distance and $Z_{[M]}$ and *x* are the charges on the metallic cation and on the $[N_2]^{x-}$ anion, respectively.

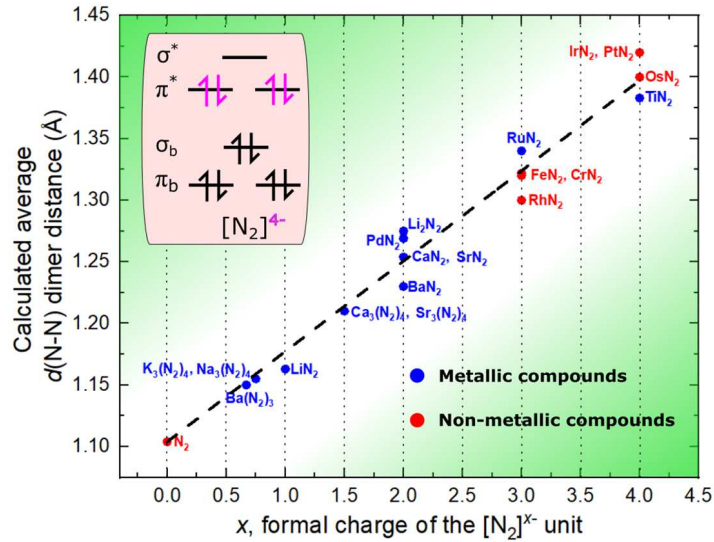

*Figure 2: Calculated (average) d(N-N) distances with respect to the formal charge x- of nitrogen dimers $[N_2]^{x-}$, for nineteen simple binary compounds containing $[N_2]^{x-}$ species.[12–20] All distances were computed at 1 bar, except for $LiN_2$ (20 GPa). The more complex binary compounds that contain $[N_2]^{x-}$ species along with another anionic nitrogen species were not included (e.g. $Sr_4N_3$,[31] $ReN_2$[36]). For the $FeN_2$,[33] $CrN_2$[34] and $RhN_2$[59] compounds, the assessment of the formal charge of the nitrogen dimer was based on their reported Raman modes and bulk moduli. Compounds featuring $[N_2]^{x-}$ species but with missing calculated data on their intramolecular bond lengths have been omitted. Since the charge on the $[N_2]^{x-}$ dimer in $CuN_2$ is not known,[12] it was not included in the plot. The linear correlation between the N-N bond lengths (BL) and the $[N_2]^{x-}$ dimers'*

*formal charge (FC) is expressed as BL = 0.074(1) Å FC + 1.104 Å and is drawn as a dashed line. The y-axis intercept at x=0 was fixed at 1.104 Å, which is the $N_2$ molecule bond length at 1 bar. Inset: Molecular orbital diagram of the pernitride $[N_2]^{4-}$.*

To understand the nature of the established correlation between the $[N_2]^{x-}$ charge and the physico-chemical properties of the *tI*44 and *cI*112 compounds, one needs to characterize the chemical bonding in these solids. Bond population analysis on all *tI*44 compounds and the *cI*112 solid (Table S6) revealed no significant electron density between $[N_2]$-$[N_2]$, M-M and $[N_2]$-M—effectively ruling out any covalent interactions between these constituents. The electron localization function (ELF) calculations (examples for $Ba(N_2)_3$ and $Sr_3(N_2)_4$ are shown in Figure 3) provide a visual corroboration. Aside from intramolecular bonding, there is no electron localization between atoms, as would have been expected for electrides. As such, then, in the first approximation, $[N_2]^{x-}$ units can be considered as isolated and their electronic structure may be illustrated with a molecular orbital diagram, like the one shown for $[N_2]^{4-}$ in the inset of Figure 2. Electrons transferred from the cations are necessarily filling the π* antibonding orbital—the lowest non-filled energy level of the $[N_2]^{x-}$ species.

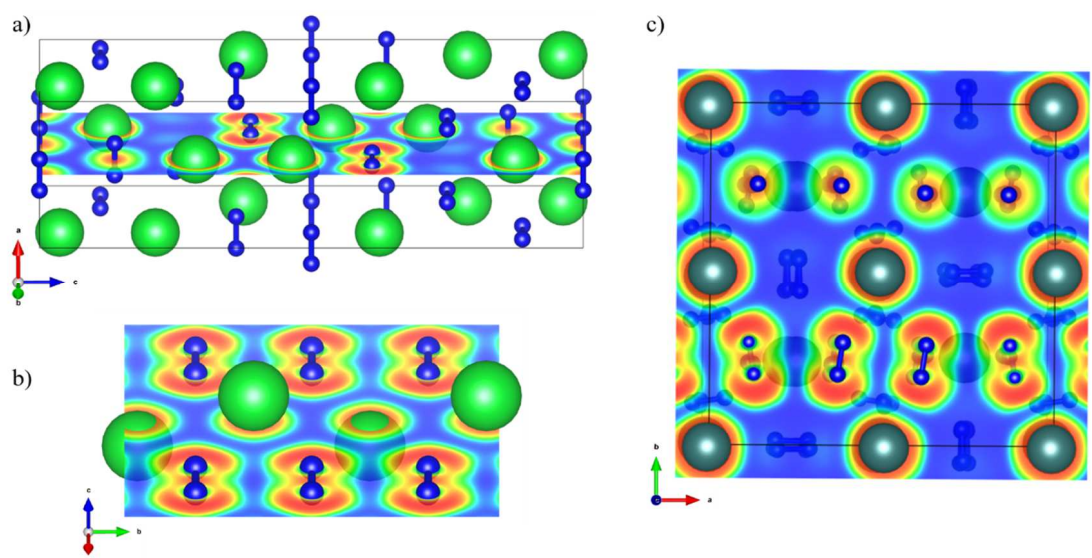

*Figure 3: Visualization of the electron localization function (ELF) analysis at 1 bar. The blue and red colors represent, respectively, no to little electron localization and a significant electron localization. The lack of pronounced electron localization in $Sr_3(N_2)_4$ between a) Sr-Sr and Sr-$[N_2]$ units ((200) slice) and b) between $[N_2]$-$[N_2]$ ((002) slice) is visible. Likewise, no notable electron localization is seen in $Ba(N_2)_3$ ((001) slice) between any species aside from the $[N_2]^{x-}$ intramolecular bond. In both compounds, electron localization is never found away from an atomic site, as it would be the case for an electride.*

The most important clue to understanding these compounds is the analysis of their electron density of states (DOS), shown in Figure 4 for $Ba(N_2)_3$ (and Figures S7-S10 for the *tI*44 solids). It reveals that *i)* all solids are metallic—agreeing well with the absence of a detected signal from Raman spectroscopy measurements made on all of them—and that *ii)* essentially all the electrons in the partially filled band are contributed by the $\pi^*$ orbitals of the charged $[N_2]^{x-}$ dimers. Indeed, the latter is verified quantitatively, where the filling of the $\pi^*$ orbitals, as reflected in the partial density of state (pDOS), was computed using the following formula:

$$\frac{\int_{-\varepsilon}^{\varepsilon_F} pDOS(E)dE}{\int_{-\varepsilon}^{+\varepsilon} pDOS(E)dE}$$

where $-\varepsilon$ denotes a suitably chosen energy level, $\varepsilon_F$ is the Fermi energy and $+\varepsilon$ is the upper energy limit. The results are shown in Table 2 and compared to the filling expected based on the formal charge. For example, in the $Ba^{2+}[(N_2)_3]^{2-}$ compound, the formal charge is -2/3 for each $[N_2]$ unit, and since four electrons can be fitted into the $\pi^*$ orbitals, their filling is expected to be of $(2/3)/4 = 0.17$. For the $Ba(N_2)_3$ compound, the $\pi^*$ orbitals' filling was found to be of 0.16 based on the integrated pDOS—closely matching the 0.17 value obtained when assuming a full cation-to-anion valence electron transfer. The same remarkable agreement is observed for all four *tI*44 compounds.

*Table 2: Comparison of the filling of the $\pi^*$ orbitals of the $[N_2]^{x-}$ dimer in the tI44 and cI112 compounds at 1 bar to the filling expected based on their formal charge. The N partial DOS was employed for these calculations (see Figures S7-S11). Essentially the same values are obtained when calculated at a pressure close to that at which the tI44 and cI112 compounds were synthesized (see Table S7).*

| Compound | $\frac{\int_{-\varepsilon}^{\varepsilon_F} pDOS(E)dE}{\int_{-\varepsilon}^{+\varepsilon} pDOS(E)dE}$ | Filling based on the formal charge |
|---|---|---|
| $Na_3(N_2)_4$ | 0.18 | 0.19 |
| $K_3(N_2)_4$ | 0.18 | 0.19 |
| $Ca_3(N_2)_4$ | 0.37 | 0.38 |
| $Sr_3(N_2)_4$ | 0.35 | 0.38 |
| $Ba(N_2)_3$ | 0.16 | 0.17 |

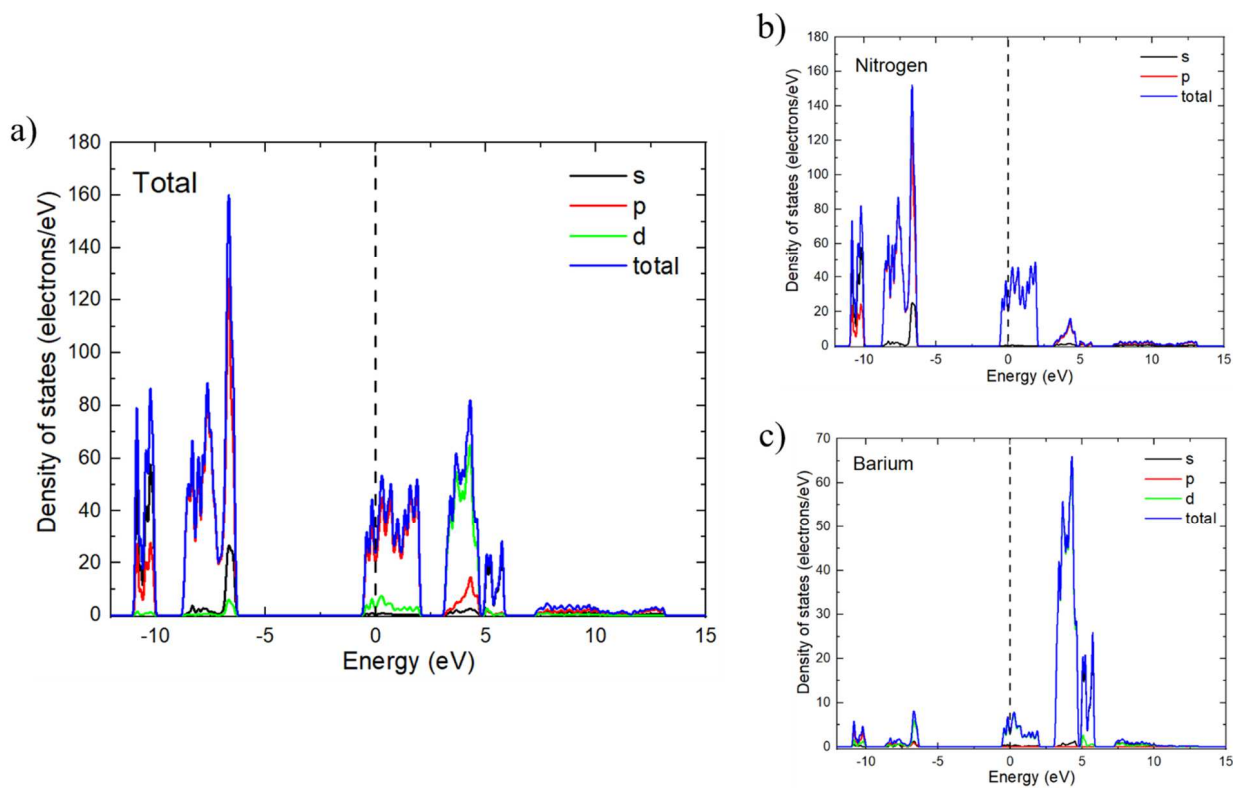

*Figure 4: Calculated electronic density of states (DOS) for the Ba(N$_2$)$_3$ compound at 1 bar. a) The total DOS with contributions from all atoms. The compound is metallic, as seen by the non-zero density of states at the Fermi energy (set at 0 eV). b) Partial DOS calculations, where only the electronic states belonging to nitrogen atoms were considered. The electrons constituting the partially filled band are lying on the π\* antibonding orbitals of the [N$_2$]$^{x-}$ species. c) Partial DOS calculations, this time only showing the electronic states formed by the barium atoms. The equivalent graphs can be found for the Na$_3$(N$_2$)$_4$, K$_3$(N$_2$)$_4$ and Sr$_3$(N$_2$)$_4$ in the supplementary figures Fig. S6-S9.*

The considerations described above enable us to conclude that (a) the valence electrons of the metal (Na, Ca, Sr, Ba) are almost entirely transferred to the nitrogen dimers, yielding cations (Na$^+$, Ca$^{2+}$, Sr$^{2+}$, Ba$^{2+}$) and [N$_2$]$^{x-}$ anions, analogously to ionic compounds; (b) the metallicity of the *tI*44 and *cI*112 solids is driven by the charged [N$_2$]$^{x-}$ anions; and (c) the conduction electrons are effectively delocalized across the π\* orbitals of all [N$_2$]$^{x-}$ units—thus responsible for the exotic non-integer charge of the dinitrogen species.

The metallicity of the *tI*44 and *cI*112 is anion-driven, making them different from common metals, alloys, metallic subnitrides (*e.g.* Ca$_2$N, Figure S12), or even metallic nitrides (*e.g.* NiN$_2$,[38] CuN$_2$,[12] CoN$_2$,[37] or TiN$_2$[20]), in which the metals (e.g. Ca, Ni, Cu, Co, Ti) contribute the totality, the majority, or a significant fraction of the electrons in the conduction band. The *tI*44 and *cI*112 solids also differ from Zintl phases, among which metallic compounds are rare and explained by a heteronuclear mixing of electronic orbitals (*i.e.* covalent bonding) leading to a partially filled band.[60,61] The defining feature, *anion-driven metallicity*—here demonstrated to impact the $[N_2]^{x-}$ units' geometry, phonon modes, and compressibility—makes the *tI*44 and *cI*112 solids representatives of a distinctive class of materials. Other synthetic substances, for example, polynitrogen compounds[8,9,62] and metal fullerides,[63] can also be considered to belong to the class of materials with *anion-driven metallicity*.

**Conclusion**

The investigation of the Na-, Ca-, Sr- and Ba-N systems up to 70 GPa has led to the synthesis of the $Na_3(N_2)_4$, $Ca_3(N_2)_4$, $Sr_3(N_2)_4$ and $Ba(N_2)_3$ compounds. In these solids, nitrogen dimers $[N_2]^{x-}$ possess non-integer formal charges $x$ equal to -0.67, -0.75 or -1.50, thus breaking the hitherto dominating paradigm of integer-charged $[N_2]^{x-}$ dimers. The values of the charges were shown to correlate with the lengths of the dimer and the physico-chemical properties of the dimer-bearing compound, thus demonstrating the fundamental importance of this parameter. The essentially complete valence electron transfer from the metal cations to the $[N_2]^{x-}$ species gives rise to anion-driven metallicity, with these electrons being delocalized on the antibonding $\pi^*$ orbitals of nitrogen dimers.

This study reveals the unique nature of the $[N_2]^{x-}$ dimers and explains the origin of their non-integer formal charges through insight into the electronic structure of the novel binary alkali- and alkaline earth metal-nitrogen compounds. These results should stimulate further analyses of the electronic structure and properties of other metallic $[N_2]^{x-}$-bearing solids. The flexibility of the $[N_2]^{x-}$ species as electron acceptors, combined with the sensitivity of their properties to the value of $x$, opens the way to electron-tunable materials and novel technologies.

**Acknowledgements**

The authors are grateful to Prof. Roald Hoffmann for fruitful discussion of the manuscript. The authors acknowledge the Deutsches Elektronen-Synchrotron (DESY, PETRA III) and the Advance Photon Source (APS) for provision of beamtime at the P02.2 and 13-IDD beamlines, respectively. D.L. thanks the Alexander von Humboldt Foundation for financial support. N.D. and L.D. thank the Federal Ministry of Education and Research, Germany (BMBF, grant no. 05K19WC1) and the Deutsche Forschungsgemeinschaft (DFG projects DU 954–11/1, DU 393–9/2, and DU 393–13/1) for financial support. B.W. gratefully acknowledges funding by the DFG in the framework of the research unit DFG FOR2125 and within projects WI1232 and thanks BIOVIA for support through the Science Ambassador program. N.D. thanks the Swedish Government Strategic Research Area in Materials Science on Functional Materials at Linköping University (Faculty Grant SFO-Mat-LiU No. 2009 00971).